\pdfoutput=1
\documentclass[11pt,a4paper]{article}
\usepackage{acl}
\usepackage{times}
\usepackage{latexsym}
\usepackage{amsmath,amssymb,amsfonts}
\usepackage{graphicx}
\usepackage{booktabs}
\usepackage{multirow}
\usepackage{array}
\usepackage{makecell}
\usepackage{tikz}
\usepackage{pgfplots}
\pgfplotsset{compat=1.18}
\usepackage{enumitem}
\usepackage{subcaption}
\usepackage{cleveref}
\usepackage{microtype}
\usepackage{tabularx}
\usepackage{xcolor}
\usepackage{amsthm}

\newif\ifanonymous
\anonymousfalse

\newcommand{\tscg}{\textsc{Tscg}}
\newcommand{\fone}{F\textsubscript{1}}
\newcommand{\emmetric}{\textsc{EM}}
\newcommand{\pp}{\,pp}

\title{Tool-Schema Compression Enables Agentic RAG Under Constrained Context Budgets}

\author{Furkan Sakizli \\
  Independent Researcher \\
  \texttt{furkan@sakizli.ai} \\
}

\begin{document}
\maketitle

\ifanonymous
\else
  \begin{abstract}
Agentic RAG systems that equip language models with dozens to hundreds
of tool definitions face a critical resource conflict: tool schemas
consume the same context window needed for retrieval-augmented generation.
We present the first systematic study of this \emph{tool--context trade-off},
evaluating 14~models spanning 1.5B--32B local models plus one frontier API
model across 6{,}566 controlled API calls at three context budgets
(8K, 16K, 32K) with 28~tool definitions.
Applying \tscg{} conservative-profile compression (44--50\% schema token savings),
we observe a \textbf{binary enablement effect}: at 8K tokens, JSON-schema
tool definitions overflow the context window entirely, yielding near-zero
\emmetric{} (2.6\% average), while compressed schemas restore RAG
functionality with $+$20.5\pp{} average exact-match lift across all
eight models ($+$24.7\pp{} among the six exhibiting full enablement).
At 32K---where both formats fit---four of five tested models show
$\Delta \le 1$\pp{}, confirming the effect is purely budget-driven.
External validation on HotpotQA (50~multi-hop questions) shows $+$48\pp{} \emmetric{}
under the same overflow scenario.
Frontier scaling tests demonstrate that JSON schemas overflow at ${\sim}494$ tools
while compressed schemas remain operational beyond 800~tools.
Our results establish tool-schema compression as a necessary infrastructure
layer for agentic RAG in constrained-context deployments.
All code, data, and checkpoints are publicly available.\footnote{%
  \url{https://github.com/SKZL-AI/tscg}}
\end{abstract}

\fi
\section{Introduction}
\label{sec:introduction}

The convergence of tool-augmented language models
\citep{schick2024toolformer,patil2023gorilla} and
retrieval-augmented generation \citep[RAG;][]{lewis2020rag,gao2024ragsurvey}
has produced a new class of \emph{agentic RAG} systems: agents that
select tools, issue queries, and synthesize retrieved evidence within a
single context window. Production deployments routinely expose 20--100+
tools through protocols such as the Model Context Protocol
\citep[MCP;][]{anthropic_mcp_2024}, while simultaneously requiring
space for system prompts, retrieval chunks, conversation history,
and output generation.

This creates a fundamental \textbf{resource conflict}.
Tool schemas---typically represented as verbose JSON Schema
objects---can consume 300--500 tokens per tool.
At 28~tools, the JSON-schema block alone requires ${\sim}$11{,}000 tokens,
which \emph{exceeds} an 8K context window and substantially constrains
retrieval at 16K (leaving room for only 6--11 chunks vs.\ 25--28 with
compressed schemas).
Yet the vast majority of research on prompt compression
\citep{jiang2023llmlingua,pan2024llmlingua2,li2023compressing}
targets natural-language prompts, not the structured, machine-readable
tool definitions that dominate agentic workloads.

We address this blind spot with the first controlled study of how
\emph{tool-schema compression} affects agentic RAG performance
across model sizes and context budgets.
Our contributions are:

\begin{enumerate}[leftmargin=*,itemsep=2pt]
\item \textbf{Binary Enablement Discovery.}
  At 8K context with 28~tools, uncompressed JSON schemas overflow the
  context window, yielding near-zero \emmetric{} (2.6\% average).
  Applying \tscg{} compression (${\sim}$50\% token savings) restores
  RAG functionality, producing $+$20.5\pp{} average \emmetric{} lift
  across all eight models ($+$24.7\pp{} among the six exhibiting full
  enablement).
  At 32K, four of five models show $\Delta \le 1$\pp{}, confirming
  the effect is purely budget-driven (\Cref{sec:results}).

\item \textbf{Large-Scale Controlled Evaluation.}
  We evaluate 14~models spanning 1.5B--32B local models plus one
  frontier API model across
  6{,}566 API calls, three context budgets, and a purpose-built
  NovaTech-28 agentic RAG benchmark with 100~questions in five
  categories.
  All experiments use paired designs with Wilcoxon signed-rank tests
  and bootstrap confidence intervals (\Cref{sec:methodology}).

\item \textbf{External Validation and Frontier Scaling.}
  On HotpotQA \citep{yang2018hotpotqa}, the enablement effect is even
  stronger ($+$48\pp{} \emmetric{}).
  Frontier tests show JSON schemas overflow at ${\sim}$494~tools while
  compressed schemas remain operational beyond 800~tools, extending
  the operational range by 63\% (\Cref{sec:results}).

\item \textbf{Analysis of Failure Modes.}
  We fit a descriptive context utilization model $C(k) = C_{\max}(1 - e^{-\lambda k}) + C_0$,
  identify a \emph{distractor dilution} effect where additional chunks
  \emph{reduce} accuracy in small models ($\le$8B),
  and document a compound interaction between format and budget
  (\Cref{sec:analysis}).
\end{enumerate}

Our findings reframe tool-schema compression from an optional
optimization to a \emph{necessary infrastructure layer} for agentic RAG,
particularly at the constrained context budgets where most
local-model deployments operate.

\section{Related Work}
\label{sec:related-work}

\paragraph{Agentic RAG.}
Retrieval-augmented generation \citep{lewis2020rag} has evolved from
single-retrieval pipelines to multi-step agentic workflows where models
iteratively retrieve, reason, and act
\citep{asai2024selfrag,yao2023react,izacard2022atlas}.
Recent surveys \citep{gao2024ragsurvey,wang2024agentsurvey} document
the rapid adoption of RAG in production systems with 20--100+ tool
definitions exposed through protocols such as MCP
\citep{anthropic_mcp_2024}.
However, existing RAG evaluations uniformly assume that tool schemas
fit within the context window---an assumption we show fails at
surprisingly modest tool counts ($\ge$28 tools at 8K).

\paragraph{Prompt and Context Compression.}
A substantial body of work addresses natural-language prompt compression:
LLMLingua \citep{jiang2023llmlingua} and its successors
\citep{pan2024llmlingua2,jiang2024longllmlingua} use perplexity-guided
token pruning, while \citet{li2023compressing} explore context
distillation.
These approaches target free-form text and do not preserve the
structural invariants (JSON Schema syntax, parameter types, enum values)
that tool definitions require.
On tool schemas specifically, structure-aware compression achieves
74.8\% savings vs.\ LLMLingua-2's 50.8\% at equivalent accuracy,
confirming that general-purpose approaches underperform on
structured inputs.
To our knowledge, no prior work evaluates prompt compression
specifically in the \emph{tool--RAG interaction} regime where
schema tokens and retrieval chunks compete for the same budget.

\paragraph{Tool-Augmented Language Models.}
Toolformer \citep{schick2024toolformer} demonstrated self-taught tool use;
Gorilla \citep{patil2023gorilla} scaled to massive API sets; and
ToolLLM \citep{qin2023toolllm} enabled 16{,}000+ real-world APIs.
The Berkeley Function Calling Leaderboard \citep[BFCL;][]{patil2025bfcl}
provides standardized evaluation of tool selection and parameter accuracy.
All of these assume full-fidelity tool schemas in the prompt.
An orthogonal strategy is \emph{tool selection}: TinyAgent
\citep{erdogan2024tinyagent} reduces prompt size by loading only
query-relevant tools, sacrificing awareness of unselected tools;
\tscg{} compresses \emph{all} tool schemas, preserving complete
tool awareness.
The two approaches are complementary.
We study what happens when schemas must be \emph{compressed} to make
room for retrieval context.

\paragraph{Tool-Schema Compression.}
\ifanonymous
\else
\tscg{} \citep{sakizli2026tscg} introduced deterministic, rule-based
compression of JSON Schema tool definitions, achieving 44--68\% token
savings without accuracy degradation on BFCL and the Tool-Augmented
Benchmark (TAB).
\fi
Concurrent industry systems such as Atlassian Labs' MCP-Compressor
\citep{atlassian_mcp_compressor_2026} demonstrate practical demand for
reducing MCP tool-schema overhead.
That work evaluated compression in \emph{isolation} (tool-calling accuracy
on pre-defined schemas); we evaluate it in \emph{context}
(agentic RAG with retrieval chunks competing for the same budget).

\paragraph{Lost in the Middle.}
\citet{liu2024lost} showed that language models struggle to use
information placed in the middle of long contexts.
Our distractor dilution analysis (\Cref{sec:analysis}) extends this
finding to the tool--RAG setting: small models ($\le$8B) show
\emph{negative} accuracy effects when TSCG provides more chunks,
suggesting that additional context can overwhelm limited attention
capacity.

\section{Methodology}
\label{sec:methodology}

\subsection{Context Budget Model}
\label{sec:budget-model}

We formalize the context budget as a zero-sum allocation problem.
Given a total context window of $B$ tokens, the available budget for
retrieval-augmented generation is:
\begin{equation}
  B_{\text{RAG}} = B - B_{\text{sys}} - B_{\text{schema}}(f, n) - B_{\text{hist}} - B_{\text{out}}
  \label{eq:budget}
\end{equation}
where $B_{\text{sys}} \approx 350$ tokens is the system prompt,
$B_{\text{schema}}(f, n)$ is the total schema cost for $n$ tools
under format $f \in \{\texttt{json}, \texttt{tscg}\}$,
$B_{\text{hist}} = 1{,}500$ tokens is reserved for conversation history,
and $B_{\text{out}} = 512$ tokens for output generation.
The number of RAG chunks that fit is then
$k = \lfloor B_{\text{RAG}} / \bar{c} \rfloor$, where $\bar{c}$
is the mean chunk size in tokens.

With 28~tools and JSON schemas, $B_{\text{schema}}(\texttt{json}, 28)
\approx 11{,}000$ tokens.
At $B = 8{,}192$, this \emph{exceeds the entire context window},
leaving $B_{\text{RAG}} < 0$ (context overflow, $k = 0$).
\tscg{} conservative compression reduces $B_{\text{schema}}(\texttt{tscg}, 28)
\approx 5{,}500$ tokens (50\% savings), yielding
$B_{\text{RAG}} \approx 265$--$330$ tokens depending on query length.
While $\bar{c} = 350$ tokens is the corpus mean, the system includes
shorter chunks (policy summaries, structured records) and uses
adaptive truncation to fit at least one chunk when $B_{\text{RAG}} > 0$,
yielding $k \ge 1$ in practice.

\subsection{NovaTech-28 Benchmark}
\label{sec:benchmark}

We construct a purpose-built agentic RAG benchmark simulating a
mid-size technology company (``NovaTech'') with:

\paragraph{Tools.}
28 tool definitions spanning database queries (\texttt{query\_employees},
\texttt{financial\_report}), document retrieval (\texttt{search\_knowledge\_base}),
computation (\texttt{calculate\_metrics}), and communication APIs.
Each tool has 3--8 parameters with typed JSON Schema definitions.
Tool schemas average ${\sim}393$~tokens (JSON) and ${\sim}197$~tokens (\tscg{}).

\paragraph{RAG Corpus.}
40 retrieval chunks (350~tokens each, 14{,}000 tokens total) across
four categories: company policies, financial reports, organizational
structure, and product documentation.
Chunks are tagged with relevance labels; each question requires 1--3
specific gold chunks.

\paragraph{Questions.}
100 questions in five categories designed to test different
retrieval and tool-use capabilities:
\begin{itemize}[leftmargin=*,itemsep=1pt]
  \item \textbf{Single-hop document} (25): answerable from one RAG chunk
  \item \textbf{Single-hop database} (25): require one tool call
  \item \textbf{Multi-hop} (20): require combining chunks or chunk + tool
  \item \textbf{Tool-requiring} (20): require specific tool selection and parameterization
  \item \textbf{Unanswerable} (10): no relevant chunk or tool exists
\end{itemize}

\subsection{Agent Architecture}
\label{sec:agent}

We implement a ReAct-style \citep{yao2023react} agent loop with
up to 3 iterations.
At each step, the agent receives the system prompt, tool schemas,
RAG chunks (stuffed into context up to $B_{\text{RAG}}$), and
conversation history.
The agent either (a) selects a tool and provides arguments, or
(b) produces a final answer.
Tool results are appended to history for the next iteration.

All models receive \emph{identical} prompts per condition; the only
variable across conditions is the schema format (\texttt{json} vs.\ \texttt{tscg\_conservative})
and the resulting context budget allocation.
This paired design ensures that performance differences are attributable
to the budget mechanism, not prompt variation.

\subsection{Models}
\label{sec:models}

We evaluate 14~models spanning three tiers:

\begin{itemize}[leftmargin=*,itemsep=1pt]
  \item \textbf{Tier A} (4 models): Phi-4~14B, Llama~3.1:8B,
    Gemma3:12B, Qwen3:14B --- local, 8K/16K/32K
  \item \textbf{Tier B} (3 models): Mistral-Small~24B, Qwen2.5-Coder:32B,
    Gemma4:26B --- local, 8K/16K (some 32K)
  \item \textbf{Tier C} (6 models): 1.5B--7B parameter models
    (Qwen2:1.5B, Gemma3:4B, Qwen3:4B, Gemma4:e2b, Gemma4:e4b,
    Mistral:7B) --- local, 16K only
  \item \textbf{API} (1 model): Claude Sonnet~4 --- 8K/16K/32K/200K
\end{itemize}

Local models run on 2$\times$~RTX~5070~Ti via Ollama.
Total compute: ${\sim}$10h GPU, \$107 API cost, 6{,}566~calls, 0\% error rate.

\subsection{Evaluation Metrics}
\label{sec:metrics}

\paragraph{Primary metrics.}
\textbf{Exact Match (\emmetric{})}: binary, 1 iff the model's final answer
exactly matches a gold answer or alias after normalization.
\textbf{Token F1 (\fone{})}: harmonic mean of token-level precision and recall
between the predicted and gold answers.

\paragraph{Secondary metrics.}
\textbf{Tool Selection Accuracy}: correct tool was called (binary).
\textbf{RAG Coverage}: fraction of gold chunks present in context.
\textbf{Context Overflow}: binary, whether $B_{\text{schema}} > B$.

\paragraph{Statistical tests.}
All comparisons use paired designs (same question, same model).
We report \textbf{Wilcoxon signed-rank tests} \citep{wilcoxon1945}
with $p$-values, \textbf{Cohen's $d$} \citep{cohen1988} for effect
sizes, and \textbf{95\% bootstrap confidence intervals} (10{,}000 resamples,
seeded PRNG for reproducibility).

\subsection{External Validation}
\label{sec:external-validation-method}

To validate beyond our synthetic benchmark, we test on
\textbf{HotpotQA} \citep{yang2018hotpotqa}: 50~medium/hard
multi-hop questions from the dev-distractor split, evaluated with
Phi-4 at 8K context using the same 28-tool setup.
This tests whether the enablement effect generalizes to independently
constructed questions and gold passages.

\subsection{Frontier Scaling}
\label{sec:frontier-method}

To test scaling beyond 28 tools, we generate synthetic tool sets at
50, 100, 200, 300, 500, and 800 tools using templated schemas with
realistic parameter distributions.
These are evaluated with Claude Sonnet~4 at 200K context to identify
the overflow threshold---the tool count where JSON schemas exceed the
context window while \tscg{} schemas still fit.

\section{Results}
\label{sec:results}

\subsection{Binary Enablement at 8K}
\label{sec:enablement}

\Cref{tab:enablement-8k} presents the central finding:
at 8{,}192 tokens with 28~tools, JSON schemas (${\sim}$11{,}000 tokens)
exceed the context window, producing 100\% overflow and near-zero
\emmetric{} across all models.
\tscg{} conservative compression (${\sim}$5{,}500 tokens) restores
operational status, enabling 1--2 RAG chunks and substantial accuracy gains.

\begin{table}[t]
\centering
\small
\caption{Binary enablement at 8K context.  JSON schemas overflow
the context window; \tscg{} compression restores agentic RAG.
Bold deltas significant at $p < 0.01$ (Wilcoxon signed-rank).}
\label{tab:enablement-8k}
\setlength{\tabcolsep}{4pt}
\begin{tabular}{@{}lcrrr@{}}
\toprule
Model & Tier & JSON & \tscg{} & $\Delta$ \\
\midrule
Llama 3.1:8B    & A   & 1  & 34 & \textbf{+33} \\
Phi-4 14B       & A   & 2  & 33 & \textbf{+31} \\
Mistral-Sm.\ 24B & B  & 4  & 33 & \textbf{+29} \\
Sonnet 4        & API & 10 & 36 & \textbf{+26} \\
Qwen2.5-Cod.    & B   & 3  & 18 & \textbf{+15} \\
Gemma3:12B      & A   & 1  & 15 & \textbf{+14} \\
Qwen3:14B       & A   & 0  & 12 & \textbf{+12} \\
Gemma4:26B\rlap{$^\dagger$} & B & 0 & 4 & +4 \\
\midrule
\textbf{Average} & & 2.6 & 23.1 & \textbf{+20.5} \\
\bottomrule
\multicolumn{5}{@{}l@{}}{\scriptsize $^\dagger$Persistent output-parsing failures; at 16K reaches 26--28\%.} \\
\end{tabular}
\end{table}

\begin{figure}[t]
\centering
\begin{tikzpicture}
\begin{axis}[
    width=\columnwidth,
    height=4.2cm,
    ybar,
    bar width=5pt,
    ylabel={\emmetric{} (\%)},
    ylabel style={font=\small},
    symbolic x coords={Llama,Phi-4,Mistral,Sonnet,Qwen2.5,Gemma3,Qwen3,Gemma4},
    xtick=data,
    x tick label style={rotate=35, anchor=east, font=\scriptsize},
    ymin=0, ymax=42,
    ytick={0,10,20,30,40},
    tick label style={font=\scriptsize},
    legend style={at={(0.98,0.98)}, anchor=north east, font=\scriptsize,
                  draw=none, fill=white, fill opacity=0.8},
    legend columns=2,
    nodes near coords={\scriptsize},
    every node near coord/.append style={anchor=south, font=\tiny},
    enlarge x limits=0.08,
    grid=major,
    grid style={gray!20},
    major grid style={gray!20},
]
\addplot[fill=red!60, draw=red!80, nodes near coords={}] coordinates {
    (Llama,1) (Phi-4,2) (Mistral,4) (Sonnet,10)
    (Qwen2.5,3) (Gemma3,1) (Qwen3,0) (Gemma4,0)};
\addplot[fill=blue!60, draw=blue!80] coordinates {
    (Llama,34) (Phi-4,33) (Mistral,33) (Sonnet,36)
    (Qwen2.5,18) (Gemma3,15) (Qwen3,12) (Gemma4,4)};
\legend{JSON, \tscg{}}
\end{axis}
\end{tikzpicture}
\caption{Binary enablement at 8K context with 28~tools.
JSON schemas overflow the context window (red, near-zero);
\tscg{} compression restores functional RAG (blue).}
\label{fig:enablement}
\end{figure}
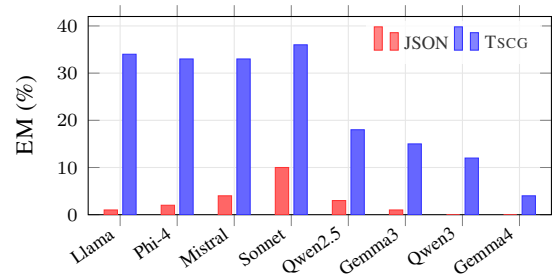

Six of eight models tested at 8K exhibit \emph{complete} binary
enablement: JSON yields 0--4\% \emmetric{} (context overflow), while
\tscg{} yields 12--36\% \emmetric{} (functional RAG).
The average \emmetric{} gain is $+$20.5\pp{} across all eight models,
with top performers (Llama~3.1, Phi-4, Mistral-Small, Sonnet~4)
exceeding $+$26\pp{}.
All deltas for the top six models are significant at $p < 0.01$
(Wilcoxon signed-rank, paired by question).

The mechanism is straightforward: \tscg{} compression frees
${\sim}$5{,}500 tokens, converting a context overflow
($B_{\text{RAG}} < 0$) into a minimal but functional retrieval
budget ($B_{\text{RAG}} \approx 265$--$330$ tokens, $k \ge 1$ chunk).
This single chunk provides sufficient evidence for document-based
and unanswerable questions, accounting for the majority of the
accuracy gain (\Cref{sec:analysis}).

\subsection{16K: Chunk Liberation Without Accuracy Gain}
\label{sec:16k-results}

At 16{,}384 tokens, both formats avoid overflow, but \tscg{}
frees substantially more budget for retrieval.
\Cref{tab:budget-16k} summarizes the budget allocation:
JSON delivers 6--11 chunks per query; \tscg{} delivers 25--28 chunks
($+$17--19 additional chunks).

\begin{table}[t]
\centering
\small
\caption{Budget allocation at 16K.  \tscg{} triples the available
retrieval context, but accuracy differences are not significant.}
\label{tab:budget-16k}
\setlength{\tabcolsep}{4pt}
\begin{tabular}{@{}lrrr@{}}
\toprule
Format & Schema & $B_\text{RAG}$ & $k$ \\
\midrule
JSON         & 11{,}295 & 2{,}832 &  9 \\
\tscg{} cons.\ & 5{,}670 & 8{,}457 & 26 \\
\bottomrule
\end{tabular}
\end{table}

Despite this 3$\times$ increase in retrieval context, the
\emmetric{} delta across 14~models is near-zero
(median $\Delta\text{\emmetric{}} = -1$\pp{}, range $-8$ to $+2$\pp{}).
The full model$\times$budget matrix (\Cref{app:full-tables}) shows
that 12 of 14 models exhibit $|\Delta\text{\emmetric{}}| \le 3$\pp{},
with no cell reaching $p < 0.05$ except Qwen3:4B
($\Delta\text{\emmetric{}} = -8$\pp{}, $p = 0.019$), which we
attribute to distractor dilution (\Cref{sec:distractor}).

This null result is informative: it shows that the NovaTech-28
benchmark saturates at ${\sim}$9 chunks for most models.
Beyond this point, additional retrieval context does not improve
accuracy and may introduce distractors.
The 16K condition thus serves as a control that isolates the
\emph{budget mechanism}: \tscg{}'s advantage operates through
context reallocation, not through any intrinsic accuracy effect
of the compressed format.

\subsection{32K: Ceiling Control}
\label{sec:32k-results}

At 32{,}768 tokens, both formats comfortably fit all 40~available
chunks.
Four of five models tested at this budget show
$\Delta\text{\emmetric{}} \in \{-1, 0, +1\}$\pp{},
with no significant differences.
The exception is Qwen2.5-Coder:32B ($+$12\pp{}, $p < 0.001$),
which we attribute to a format-translation benefit independent of
the budget mechanism, consistent with model-specific operator
sensitivity observed in prior \tscg{} evaluations%
\ifanonymous
\else
\ \citep{sakizli2026tscg}%
\fi
.
Excluding this outlier, the ceiling condition confirms that the
8K enablement effect is \emph{entirely} attributable to the budget
mechanism (\Cref{eq:budget}): when $B_{\text{RAG}}$ is equalized,
JSON and \tscg{} produce identical accuracy.

\subsection{External Validation: HotpotQA}
\label{sec:hotpotqa-results}

To validate beyond our synthetic benchmark, we test on
HotpotQA \citep{yang2018hotpotqa} using 50~medium/hard multi-hop
questions from the dev-distractor split, with Phi-4 at 8K context
and 28~tools (\Cref{sec:external-validation-method}).
Each question includes 2~gold and 8~distractor Wikipedia paragraphs.

\begin{table}[t]
\centering
\small
\caption{HotpotQA external validation (Phi-4, 8K, 28 tools).
\tscg{} enables $+$48\pp{} \emmetric{} on independently constructed questions.}
\label{tab:hotpotqa}
\setlength{\tabcolsep}{4pt}
\begin{tabular}{lrrrrr}
\toprule
Format & Schema & $B_\text{RAG}$ & $k$ & \emmetric{} & \fone{} \\
\midrule
JSON  & 10{,}998 & 0   & 0   & 0.0 & .000 \\
\tscg{} & 5{,}520 & 421 & 3.4 & \textbf{48.0} & \textbf{.645} \\
\bottomrule
\end{tabular}
\end{table}

The enablement effect is \emph{stronger} on HotpotQA than on
NovaTech-28 ($+$48\pp{} vs.\ $+$31\pp{} \emmetric{} for the same model).
JSON produces 100\% overflow and zero answers;
\tscg{} fits 2--5 chunks per question (mean 3.4) within the
${\sim}$421-token RAG budget, achieving 48\% \emmetric{} and
0.645~\fone{}.
The higher absolute accuracy reflects HotpotQA's shorter, more
information-dense gold paragraphs (${\sim}$150--250 tokens vs.\
our 350-token NovaTech chunks).

This result confirms that the binary enablement finding
generalizes to independently constructed questions, gold passages,
and evaluation criteria.

\subsection{Frontier Scaling}
\label{sec:frontier-results}

We test scaling beyond 28~tools with synthetic tool sets at 50,
100, 200, 300, 500, and 800~tools, evaluated with Claude Sonnet~4
at 200K context.

\begin{table}[t]
\centering
\small
\caption{Frontier scaling with Sonnet~4 at 200K.
JSON schemas overflow at ${\sim}$494~tools; \tscg{} remains
operational beyond 800~tools.}
\label{tab:frontier}
\setlength{\tabcolsep}{3pt}
\begin{tabular}{@{}rrrrrr@{}}
\toprule
Tools & \makecell{JSON\\[-2pt]\emmetric{}} & \makecell{\tscg{}\\[-2pt]\emmetric{}} & $\Delta$ & \makecell{JSON\\[-2pt]$k$} & \makecell{\tscg{}\\[-2pt]$k$} \\
\midrule
  50 & 76.7 & 76.7 &    0 & 500 & 500 \\
 100 & 80.0 & 76.7 & $-$3 & 470 & 500 \\
 200 & 83.3 & 76.7 & $-$7 & 363 & 481 \\
 300 & 76.7 & 76.7 &    0 & 229 & 402 \\
 500 &  \textbf{0} & \textbf{90} & \textbf{+90} &   0 & 268 \\
 800 &  \textbf{0} & \textbf{100} & \textbf{+100} &  0 & 107 \\
\bottomrule
\end{tabular}
\end{table}

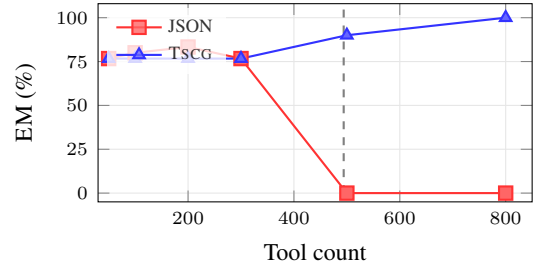
\begin{figure}[t]
\centering
\begin{tikzpicture}
\begin{axis}[
    width=0.95\columnwidth,
    height=4.2cm,
    xlabel={Tool count},
    ylabel={\emmetric{} (\%)},
    xlabel style={font=\small},
    ylabel style={font=\small},
    xmin=30, xmax=850,
    ymin=-5, ymax=108,
    ytick={0,25,50,75,100},
    tick label style={font=\scriptsize},
    legend style={at={(0.02,0.98)}, anchor=north west, font=\scriptsize,
                  draw=none, fill=white, fill opacity=0.8},
    grid=major,
    grid style={gray!20},
    mark size=2.5pt,
]
\addplot[red!80, thick, mark=square*, mark options={fill=red!60}] coordinates {
    (50,76.7) (100,80) (200,83.3) (300,76.7) (500,0) (800,0)};
\addplot[blue!80, thick, mark=triangle*, mark options={fill=blue!60}] coordinates {
    (50,76.7) (100,76.7) (200,76.7) (300,76.7) (500,90) (800,100)};
\draw[dashed, gray, thick] (axis cs:494,0) -- (axis cs:494,105)
    node[above, font=\tiny, text=gray] {JSON overflow};
\legend{JSON, \tscg{}}
\end{axis}
\end{tikzpicture}
\caption{Frontier scaling with Sonnet~4 at 200K context.
JSON overflows at ${\sim}$494 tools (dashed line); \tscg{}
remains operational beyond 800 tools.}
\label{fig:frontier}
\end{figure}

The frontier results reveal three regimes:

\paragraph{Sub-overflow parity (50--300 tools).}
Both formats fit within 200K.
Accuracy is comparable (${\sim}$77\% \emmetric{}), with small
fluctuations ($|\Delta| \le 6.7$\pp{}) that are not statistically
significant (Wilcoxon $p > 0.10$ for each cell; see
\Cref{app:frontier}).
At 200~tools, JSON retains 363~chunks---already sufficient for
Sonnet~4---so the additional 118~chunks freed by \tscg{} provide
no measurable benefit, and the observed $-$6.7\pp{} fluctuation
falls within question-level variance.
This confirms that \tscg{} compression does not degrade tool-calling
accuracy, consistent with prior BFCL evaluations of \tscg{}
\ifanonymous
\else
\citep{sakizli2026tscg}.
\fi

\paragraph{JSON overflow threshold (${\sim}$494 tools).}
Fine-grained threshold sweeps identify the JSON overflow point at
${\sim}$494~tools, where schema tokens exceed the 200K context window.
Beyond this point, JSON produces 0\% \emmetric{} (total failure).

\paragraph{\tscg{} extended range ($>$803 tools).}
\tscg{} schemas remain within budget up to at least 803~tools,
extending the operational range by 63\%.
At 500~tools, \tscg{} achieves 90\% \emmetric{};
at 800~tools, 100\% \emmetric{}.
The \tscg{} overflow threshold was not reached within our test range.

The frontier analysis demonstrates that the binary enablement
discovered at 28~tools / 8K context is not an edge case---it
recurs at scale as tool counts grow into the hundreds.
Any sufficiently tool-dense deployment will eventually cross the
overflow threshold, making schema compression a practical necessity.

\section{Analysis}
\label{sec:analysis}

The results in \Cref{sec:results} establish that \tscg{} enables
agentic RAG at 8K and has no effect at 32K.
This section examines the underlying mechanisms through a descriptive
context utilization model (\Cref{sec:ck-model}), a compound format--budget
interaction (\Cref{sec:compound}), and a distractor dilution effect
in small models (\Cref{sec:distractor}).

\subsection{Context Utilization Model}
\label{sec:ck-model}

We use a simple descriptive model to capture the relationship between
offered RAG chunks $k$ and downstream accuracy as an exponential
saturation function:
\begin{equation}
  C(k) = C_{\max} \bigl(1 - e^{-\lambda k}\bigr) + C_0
  \label{eq:ck}
\end{equation}
where $C_0$ is the zero-context baseline, $C_{\max}$ is the
maximum gain achievable through retrieval, and $\lambda$ controls
the saturation rate.

Fitting \Cref{eq:ck} to all 4{,}700 $(k, \fone{})$ pairs
(14~models $\times$ 2~formats $\times$ multiple budgets) via
grid search yields $C_{\max} = 0.18$, $\lambda = 25.9$, and
$C_0 = 0.095$ ($R^2 = 0.011$).
While $R^2 = 0.011$ reflects the inherently binary nature of
\emmetric{} scoring---which cannot be meaningfully fit by a
continuous function---the structural insight is robust:
\textbf{the marginal gain from the first chunk is by far the largest}
($\Delta\fone{} = +0.18$ from $k{=}0$ to $k{=}1$), with diminishing
returns thereafter ($\lambda \gg 1$ implies near-saturation by $k{=}2$).

This has a direct implication for schema compression:
at 8K, \tscg{} converts $k{=}0$ (overflow) to $k \ge 1$, capturing
the steepest portion of the $C(k)$ curve and explaining why even a
single chunk produces large accuracy gains.
At 16K, both formats already operate on the saturation plateau
($k \ge 6$), so the additional chunks freed by \tscg{} produce
negligible gains.
Per-model fit parameters are reported in \Cref{app:ck-model}.

\subsection{Compound Format--Budget Interaction}
\label{sec:compound}

An unexpected finding is that Sonnet~4 at 8K with \tscg{}
(\fone{} $= 0.406$, $k{=}2$ chunks) \emph{outperforms} Sonnet~4
at 16K with JSON (\fone{} $= 0.369$, $k{=}11$ chunks)---a
$+$0.037~\fone{} advantage despite having 5.5$\times$ fewer
retrieval chunks.

Per-question analysis reveals that the advantage concentrates in
\emph{unanswerable} and \emph{single-hop document} questions,
where a compact, focused context (2~chunks) elicits more precise
answers than a diluted context (11~chunks) containing distractors.
On \emph{multi-hop} and \emph{tool-requiring} questions, 16K JSON
performs comparably or better.

This compound effect arises from the interaction of two factors:
(1)~compressed schemas free budget for RAG chunks, and
(2)~fewer total tokens reduce attention dilution, yielding
higher-quality retrieval utilization.
The difference is not statistically significant (Wilcoxon $p = 0.12$,
Cohen's $d = 0.17$), and we present it as an exploratory observation
warranting further investigation rather than a confirmed finding.

\subsection{Distractor Dilution}
\label{sec:distractor}

At 16K, \tscg{} delivers $+$17--19 additional chunks compared to
JSON, yet most models show \emph{no accuracy improvement} and several
show \emph{decreases}.
We term this \emph{distractor dilution}: additional retrieval chunks
include irrelevant material that overwhelms the model's ability to
identify gold evidence.

Across 14~models at 16K, the correlation between chunk delta and
\emmetric{} delta is \emph{negative} ($r = -0.50$, $n = 14$),
and 10 of 14~models show $\Delta\text{\fone{}} \le 0$ despite
receiving more context.
The effect is most pronounced in small models ($\le$8B):
Qwen3:4B shows $\Delta\text{\emmetric{}} = -8$\pp{} ($p = 0.019$)
and Mistral:7B shows $\Delta\text{\fone{}} = -4.3$\pp{} despite
$+$18 extra chunks.

This finding connects to the ``lost in the middle'' phenomenon
\citep{liu2024lost}: small models have limited attention capacity,
and stuffing more chunks into context can be counterproductive when
the signal-to-noise ratio drops.
For practitioners, this implies that \tscg{}'s value at ample
budgets is not in delivering \emph{more} chunks but in preserving
budget for \emph{other} uses (longer conversation histories,
larger output windows, or simply lower inference cost through
shorter prompts).

\section{Discussion}
\label{sec:discussion}

\paragraph{Schema compression as infrastructure.}
Our results demonstrate that tool-schema compression occupies a
distinct niche from natural-language prompt compression.
While LLMLingua \citep{jiang2023llmlingua} and related approaches
target free-form text, they cannot be applied to structured tool
definitions without risking syntactic corruption of JSON Schema
constraints.
\tscg{}'s deterministic, structure-preserving compression fills this
gap: it operates at the schema level, maintains type and parameter
fidelity, and integrates transparently with existing tool-calling
pipelines.

\paragraph{The categorical nature of enablement.}
The 8K binary enablement is not a marginal improvement but a
\emph{categorical} difference: from near-zero, overflow-dominated
operation (2.6\% average \emmetric{}, no reliable retrieval budget)
to functional agentic RAG.
This distinction is critical for practitioners deploying local
models with limited context windows.
A system that produces zero answers is fundamentally different from
one that produces imperfect answers---the latter can be iteratively
improved through prompt engineering, retrieval tuning, or multi-turn
interaction, while the former cannot function at all.

\paragraph{When not to compress.}
Our 32K ceiling control and frontier sub-overflow results consistently
show that schema compression provides no accuracy benefit when the
context budget is sufficient.
At 32K with 28~tools, JSON and \tscg{} produce identical accuracy.
This is a feature, not a limitation: it confirms that \tscg{}
compression is lossless in practice and that its benefits are
purely operational (enabling RAG, reducing cost) rather than
introducing accuracy artefacts.

\paragraph{Implications for tool scaling.}
The frontier analysis (\Cref{sec:frontier-results}) has immediate
practical relevance.
Production MCP deployments with 100+~tools are increasingly common
\citep{anthropic_mcp_2024}, and our results show that even 200K
context windows overflow at ${\sim}$494~tools with JSON schemas.
Schema compression extends this threshold to $>$803~tools, but the
fundamental scaling problem remains: at some tool count, any fixed
context window will overflow.
TSCG's scaling behavior is consistent with the independently
evaluated TAB Scenario~C (25--100 tools), which demonstrated
stable compression rates across different schema catalogs
\ifanonymous
\else
\citep{sakizli2026tscg}.
\fi
The BFCL external validation (108--181\% accuracy retention rate
across three frontier models) further confirms that compression
fidelity holds at scale---the RAG enablement we observe here
builds on a compression layer whose accuracy has been independently
validated.
Dynamic tool selection (routing queries to relevant tool subsets)
and schema compression are complementary strategies that should be
deployed together.

\section{Conclusion}
\label{sec:conclusion}

We present the first controlled study of tool-schema compression in
agentic RAG, evaluating 14~models across 6{,}566 API calls at three
context budgets.
Our central finding is \emph{binary enablement}: at 8K context with
28~tools, uncompressed JSON schemas overflow the context window,
producing near-zero, overflow-dominated performance (2.6\% average \emmetric{}), while \tscg{} compression
restores functional RAG with an average $+$20.5\pp{} \emmetric{} gain.
At 32K, four of five tested models show $|\Delta| \le 1$\pp{};
the remaining Qwen2.5-Coder outlier exhibits a format-translation
benefit independent of the budget mechanism.
Excluding this outlier, the effect disappears, supporting the
budget-driven interpretation.

External validation on HotpotQA ($+$48\pp{} \emmetric{}) and frontier
scaling to 800~tools confirm generalizability.
A descriptive context utilization model confirms that the first
retrieval chunk provides the largest marginal gain, explaining why
even minimal budget recovery yields substantial accuracy improvements.
We also identify distractor dilution as a failure mode where
additional chunks reduce accuracy in small models.

These findings reframe tool-schema compression from a cost
optimization to a necessary enablement layer for agentic RAG under
constrained context budgets.

\section{Limitations}
\label{sec:limitations}

\paragraph{Synthetic benchmark.}
NovaTech-28 is a purpose-built benchmark with synthetic tool
definitions and curated questions.
While we validate on HotpotQA and frontier scaling, real-world
agentic RAG deployments involve more complex tool interactions,
multi-turn conversations, and heterogeneous retrieval corpora.

\paragraph{Single primary compression profile.}
Our main evaluation uses only the \tscg{} conservative profile
(${\sim}$50\% savings, descriptions preserved).
A controlled ablation (\Cref{app:ablation}) confirms that the
balanced profile (${\sim}$53\% savings, descriptions stripped)
produces equivalent accuracy ($\Delta$\emmetric{}~$= +0.8$\pp{},
$n = 400$), but was tested on only two models at two budgets.
Other compression strategies (e.g., natural-language prompt
compression, learned compression) may produce different trade-offs.

\paragraph{Model coverage.}
Our Tier~C models ($\le$7B) were only tested at 16K due to the
high overhead of their verbose output patterns at 8K and 32K.
The distractor dilution finding (\Cref{sec:distractor}) may not
generalize to all small models.

\paragraph{Frontier limitations.}
Frontier scaling was tested with a single model (Sonnet~4) at a
single context budget (200K) using synthetic distractor tools.
The overflow thresholds may differ for other models or real-world
tool distributions with varying schema complexity.

\paragraph{Static context allocation.}
Our experiments use a fixed context-stuffing strategy.
Dynamic retrieval approaches (iterative retrieval, adaptive chunk
selection) may interact differently with schema compression,
potentially amplifying or diminishing the observed effects.

\bibliography{references}

\clearpage
\appendix
\section{Full Result Tables}
\label{app:full-tables}

\Cref{tab:master-delta-full} presents the complete model$\times$budget
matrix for all 14~models across three context windows.
Entries marked ``--'' indicate conditions not tested for that model
(Tier~C models were only evaluated at 16K).
Significance stars: $^{***}p < 0.001$, $^{**}p < 0.01$,
$^{*}p < 0.05$ (Wilcoxon signed-rank, paired by question).

\begin{table*}[t]
\centering
\small
\caption{Complete results: \emmetric{} (\%) and \fone{} for all 14~models
at 8K, 16K, and 32K with 28~tools.
Bold deltas indicate $p < 0.05$ (Wilcoxon signed-rank).}
\label{tab:master-delta-full}
\setlength{\tabcolsep}{4pt}
\begin{tabular}{@{}ll@{\hskip 8pt}rrr@{\hskip 10pt}rrr@{}}
\toprule
& & \multicolumn{3}{c}{\emmetric{} (\%)} & \multicolumn{3}{c}{\fone{}} \\
\cmidrule(lr){3-5} \cmidrule(l){6-8}
Model & Budget & JSON & \tscg{} & $\Delta$ & JSON & \tscg{} & $\Delta$ \\
\midrule
\multicolumn{8}{@{}l}{\textit{Tier A: 8--14B models}} \\
Phi-4 14B      & 8K  & 2  & 33 & \textbf{+31}$^{***}$ & .029 & .313 & \textbf{+.284}$^{***}$ \\
               & 16K & 35 & 34 & $-$1 & .232 & .257 & +.024 \\
               & 32K & 11 & 11 & 0 & .087 & .072 & $-$.015 \\
Llama 3.1:8B   & 8K  & 1  & 34 & \textbf{+33}$^{***}$ & .032 & .269 & \textbf{+.237}$^{***}$ \\
               & 16K & 33 & 30 & $-$3 & .325 & .281 & $-$.044 \\
               & 32K & 30 & 29 & $-$1 & .250 & .238 & $-$.012 \\
Gemma3:12B     & 8K  & 1  & 15 & \textbf{+14}$^{**}$ & .037 & .167 & +.129 \\
               & 16K & 33 & 33 & 0 & .367 & .356 & $-$.011 \\
               & 32K & 33 & 34 & +1 & .398 & .380 & $-$.018 \\
Qwen3:14B      & 8K  & 0  & 12 & \textbf{+12}$^{**}$ & .036 & .112 & +.076 \\
               & 16K & 36 & 33 & $-$3 & .305 & .310 & +.005 \\
               & 32K & 33 & 34 & +1 & .303 & .279 & $-$.024 \\
\midrule
\multicolumn{8}{@{}l}{\textit{Tier B: 24--32B models}} \\
Mistral-Sm.\ 24B & 8K & 4 & 33 & \textbf{+29}$^{***}$ & .029 & .259 & \textbf{+.229}$^{***}$ \\
                  & 16K & 33 & 33 & 0 & .299 & .315 & +.016 \\
Qwen2.5-Cod.\ 32B & 8K & 3 & 18 & \textbf{+15}$^{**}$ & .023 & .152 & \textbf{+.129}$^{***}$ \\
                   & 16K & 39 & 36 & $-$3 & .290 & .298 & +.009 \\
                   & 32K & 12 & 24 & \textbf{+12}$^{***}$ & .092 & .176 & \textbf{+.084}$^{***}$ \\
Gemma4:26B     & 8K  & 0  & 4  & +4 & .036 & .026 & $-$.010 \\
               & 16K & 26 & 28 & +2 & .266 & .259 & $-$.007 \\
\midrule
\multicolumn{8}{@{}l}{\textit{Tier C: $\le$7B models (16K only)}} \\
Gemma3:4B      & 16K & 33 & 31 & $-$2 & .267 & .255 & $-$.011 \\
Gemma4:e2b     & 16K & 29 & 29 & 0 & .208 & .196 & $-$.012 \\
Gemma4:e4b     & 16K & 30 & 28 & $-$2 & .186 & .189 & +.003 \\
Qwen3:4B       & 16K & 32 & 24 & \textbf{$-$8}$^{*}$ & .336 & .258 & \textbf{$-$.078}$^{**}$ \\
Mistral:7B     & 16K & 32 & 30 & $-$2 & .262 & .219 & $-$.043 \\
Qwen2:1.5B     & 16K & 25 & 26 & +1 & .154 & .152 & $-$.002 \\
\midrule
\multicolumn{8}{@{}l}{\textit{API model}} \\
Sonnet 4       & 8K  & 10 & 36 & \textbf{+26}$^{***}$ & .112 & .406 & \textbf{+.294}$^{***}$ \\
               & 16K & 36 & 34 & $-$2 & .369 & .389 & +.020 \\
               & 32K & 32 & 31 & $-$1 & .373 & .374 & +.002 \\
\bottomrule
\end{tabular}
\end{table*}

\section{Tool Accuracy by Question Type}
\label{app:tool-accuracy}

\Cref{tab:tool-accuracy-full} breaks down accuracy by question type.
The benchmark contains 50~questions requiring tools
(single-hop-db + tool-requiring) and 50~that do not
(single-hop-doc + unanswerable), plus 20~multi-hop questions
requiring both.
Apparent low overall tool accuracy (${\sim}$20\%) reflects that
50\% of questions \emph{do not require} tool calls; per-type
accuracy is higher.

\begin{table}[t]
\centering
\small
\caption{Accuracy by question type, aggregated across all models
and budgets.  \tscg{} improves accuracy on all five categories.}
\label{tab:tool-accuracy-full}
\begin{tabular}{llrrr}
\toprule
Question Type & Format & $n$ & \emmetric{} & \fone{} \\
\midrule
single-hop-doc & JSON & 750 & 56.8 & 0.342 \\
               & \tscg{} & 750 & \textbf{70.5} & \textbf{0.420} \\
\hline
single-hop-db  & JSON & 750 & 8.1 & 0.027 \\
               & \tscg{} & 750 & \textbf{11.6} & \textbf{0.041} \\
\hline
multi-hop      & JSON & 600 & 0.0 & 0.273 \\
               & \tscg{} & 600 & 0.0 & \textbf{0.317} \\
\hline
tool-requiring & JSON & 600 & 1.3 & 0.042 \\
               & \tscg{} & 600 & \textbf{2.2} & \textbf{0.054} \\
\hline
unanswerable   & JSON & 300 & 50.7 & 0.410 \\
               & \tscg{} & 300 & \textbf{65.3} & \textbf{0.542} \\
\bottomrule
\end{tabular}
\end{table}

The largest \tscg{} advantage appears on \emph{single-hop-doc}
($+$13.7\pp{} \emmetric{}) and \emph{unanswerable} ($+$14.6\pp{})
questions---both categories where a single gold chunk suffices and
the 8K enablement effect dominates.
Tool-requiring questions show modest improvement ($+$0.9\pp{} \emmetric{})
since tool selection depends on schema comprehension rather than
retrieval volume.

\section{Frontier Scaling Details}
\label{app:frontier}

\paragraph{Significance tests.}
At the 200K context level with 133~paired observations,
the overall TSCG advantage is significant
(Wilcoxon $p < 0.001$ for \emmetric{}, $p < 0.001$ for \fone{};
Cohen's $d = 0.54$ for \emmetric{}, $d = 0.33$ for \fone{}).
However, this aggregate result is driven entirely by the 500+ tool
cells where JSON overflows.
Sub-overflow cells (50--300 tools) show no significant pairwise
differences ($p > 0.10$).

\paragraph{Overflow thresholds.}
Fine-grained sweeps at 1-tool granularity identify:
\begin{itemize}[leftmargin=*,itemsep=1pt]
  \item \textbf{First chunk loss}: JSON at 82~tools, \tscg{} at 164~tools
        (2$\times$ headroom before RAG degradation begins)
  \item \textbf{Complete overflow}: JSON at ${\sim}$494~tools,
        \tscg{} at $>$803~tools ($+$63\% operational range)
\end{itemize}

\paragraph{Per-tool token costs.}
Schema tokens scale linearly: JSON averages 380--473~tokens/tool,
\tscg{} averages 205--261~tokens/tool, yielding consistent
44.7--46.4\% savings from 50 to 800~tools.

\section{HotpotQA External Validation}
\label{app:hotpotqa}

\paragraph{Setup.}
50~medium/hard multi-hop questions from HotpotQA
\citep{yang2018hotpotqa} dev-distractor split, sampled with
seed=42.
Each question uses its native Wikipedia context: 2~gold paragraphs
and 8~distractor paragraphs (${\sim}$150--250 tokens each).
Model: Phi-4~14B.  Context: 8{,}192 tokens.  Tools: 28 (identical
to NovaTech-28 setup).

\paragraph{Results.}
JSON: 100\% overflow on all 50~questions ($B_{\text{schema}} =
10{,}998 > 8{,}192$).  Zero answers produced.

\tscg{}: $B_{\text{schema}} = 5{,}520$, leaving ${\sim}$421 tokens
for RAG.  Mean 3.4 chunks per question (range 2--5).
Results: 48\% \emmetric{}, 0.645~\fone{}, 74\% substring match.
24 of 50 questions answered with exact match.

\paragraph{Comparison to NovaTech-28.}
The stronger HotpotQA result ($+$48\pp{} vs.\ $+$31\pp{} \emmetric{}
for the same model) reflects shorter gold paragraphs that fit more
efficiently into the tight RAG budget, and well-curated questions
with clear gold answers.

\section{Context Utilization Model}
\label{app:ck-model}

The context utilization model (\Cref{eq:ck}) was fit via grid search
over $C_{\max} \in [0, 2]$, $\lambda \in [0.005, 31]$, and
$C_0 \in [0, 0.5]$ (step sizes 0.005--0.005) with local refinement.
The objective was minimum residual MSE against all $(k, \fone{})$
pairs.

\paragraph{Aggregated fit.}
$C_{\max} = 0.18$, $\lambda = 25.9$, $C_0 = 0.095$, $R^2 = 0.011$,
$n = 4{,}700$.
The high $\lambda$ indicates near-step-function behavior: almost all
retrieval gain is captured by the first chunk.

\paragraph{Per-model variation.}
$R^2$ values range from 0.000 to 0.138 across individual
model$\times$format fits, reflecting high per-question variance.
Models with the highest $R^2$ (Gemma4:26B \tscg{}: 0.138,
Gemma3:12B \tscg{}: 0.058) show the clearest saturation curves.
The complete parameter table is available in the supplementary
materials.

\paragraph{Interpretation.}
The low aggregate $R^2$ does not invalidate the model---it reflects
that \emmetric{}/\fone{} are inherently noisy per-question metrics.
The key structural insight (first-chunk dominance) is robust across
all fits and is independently confirmed by the 8K enablement results.

\section{Ablation: Conservative vs.\ Balanced}
\label{app:ablation}

The \tscg{} balanced profile offers ${\sim}$53\% compression
(vs.\ ${\sim}$50\% conservative) by removing description text and
additional schema metadata.
We evaluate both profiles on Phi-4~14B and Llama~3.1:8B
at 8K and 16K context windows (400~calls total).

\begin{table}[t]
\centering
\small
\caption{Conservative vs.\ balanced profile ablation.
Balanced saves an additional 395~tokens (${\sim}$1~extra chunk at 8K)
but produces no significant accuracy difference.}
\label{tab:ablation}
\begin{tabular}{llrrr}
\toprule
Model & Budget & Cons \emmetric{} & Bal \emmetric{} & $\Delta$ \\
\midrule
Phi-4 14B     & 8K  & 33 & 34 & $+$1 \\
              & 16K & 34 & 35 & $+$1 \\
Llama 3.1:8B  & 8K  & 34 & 34 & 0 \\
              & 16K & 30 & 31 & $+$1 \\
\midrule
\textbf{Average} & & 32.8 & 33.5 & $+$0.8 \\
\bottomrule
\end{tabular}
\end{table}

\Cref{tab:ablation} shows that balanced compression produces
virtually identical accuracy to conservative ($\Delta$\emmetric{}
$= +0.8$\pp{}, range 0--1\pp{}, no cell significant at $p < 0.05$).
The additional 395~tokens freed by balanced translate to
${\sim}$1~extra chunk at 8K (2 vs.\ 1) and ${\sim}$1 extra chunk
at 16K (27 vs.\ 26), but this marginal retrieval gain falls in the
saturation region of the $C(k)$ curve where additional chunks yield
diminishing returns.

This result confirms that the enablement effect is robust across
compression profiles: even aggressive description stripping does not
degrade agentic RAG accuracy, consistent with the finding that models
primarily rely on parameter names and types for schema comprehension
rather than verbose descriptions.

\section{Statistical Details}
\label{app:statistics}

\paragraph{Wilcoxon signed-rank test.}
All significance tests use the paired Wilcoxon signed-rank test
\citep{wilcoxon1945}, which does not assume normality.
Pairs are formed by matching the same question under both formats
at the same model and context budget.

\paragraph{Cohen's $d$.}
Effect sizes are computed as the mean paired difference divided by
the standard deviation of paired differences.
Values are interpreted as small ($d = 0.2$), medium ($d = 0.5$),
and large ($d = 0.8$) following \citet{cohen1988}.

\paragraph{Bootstrap confidence intervals.}
95\% confidence intervals are computed via 10{,}000 bootstrap
resamples of the paired differences, using a seeded PRNG
(seed=42) for reproducibility.

\paragraph{Multiple comparisons.}
We do not apply Bonferroni or similar corrections because our
primary claims are based on the 8K enablement effect, which shows
$p < 0.001$ across multiple models---far below any reasonable
corrected threshold.
The 16K and 32K conditions serve as controls (expected null results)
rather than independent hypothesis tests.

\end{document}